\documentclass[]{jfm}
\usepackage{bm} 
\usepackage[normalem]{ulem} 
\usepackage{graphicx}
\usepackage{newtxmath}
\usepackage{natbib}
\usepackage{hyperref}
\usepackage{multirow}
\hypersetup{
    colorlinks = false,
    linkcolor  = blue,
    urlcolor   = blue,
    citecolor  = blue,
}

\newcommand{\RomanNumeralCaps}[1]
\linenumbers

\title[Attenuation of turbulence by spherical particles]
{Attenuation of turbulence in a periodic cube \\
by finite-size spherical solid particles}
\author[S. Oka and S. Goto]
{Sunao Oka \corresp{\email{s\_oka@fm.me.es.osaka-u.ac.jp}}
and 
Susumu Goto \corresp{\email{s.goto.es@osaka-u.ac.jp}}}

\affiliation{Graduate School of Engineering Science, Osaka University,\\
1-3 Machikaneyama, Toyonaka, Osaka, 560-8531 Japan}

\begin{document}
\maketitle

\begin{abstract}
To investigate the attenuation of turbulence in a periodic cube due to the addition of spherical solid particles, we conduct direct numerical simulations using an immersed boundary method with resolving flow around each particle. Numerical results with systematically changing particle diameters and Stokes numbers for a fixed volume fraction $\Lambda$ show that the additional energy dissipation rate in the wake of particles determines the degree of the attenuation of turbulent kinetic energy. On the basis of this observation, we propose the formulae describing the condition and degree of the attenuation of turbulence intensity. We conclude that particles with the size proportional to $\lambda/\sqrt{\gamma}$, where $\lambda$ and $\gamma$ are the Taylor length and the mass density ratio between particles and fluid, most significantly reduce the intensity of developed turbulence under the condition that $\gamma$ and $\Lambda$ are fixed.
\end{abstract}



\section{Introduction}

We investigate solid particle suspension, where flow advects particles and vortices shedding from particles can change surrounding flow. Such fluid-particle interactions play essential roles in many flow systems. In particular, the enhancement and attenuation of turbulence by the addition of solid particles are important in industrial and environmental flows. However, there remain many unsolved scientific issues on the complex phenomena. In fact, although turbulence modulation due to solid particles is a classical issue in fluid mechanics back to the seminal experiments by \citet{Tsuji-1982} and \citet{Tsuji-1984} about 40 years ago, there is no clear conclusion even for the most fundamental question: i.e.~what determines the condition for the turbulence modulation? \citet{Gore-1989} proposed a criterion on this issue by compiling the data of particulate turbulent pipe flow and jet. They concluded that turbulence was enhanced (or attenuated) if the ratio $D/L$, with $D$ and $L$ being the particle diameter and the integral length [$L=0.2\times\text{(pipe radius)}$ for pipe flow and $0.039\times\text{(distance from the exit)}$ for jet], is larger (or smaller) than $0.1$ because larger particles produce turbulence in their wake, while smaller ones acquire their energy from large-scale vortices. Since then, even recently experiments with newer techniques such as particle tracking \citep{Cisse-2015} and particle image velocimetry \citep{Hoque-2016} were conducted. However, \citet{Gore-1989}'s picture still holds, although \citet{Hoque-2016}, for example, proposed a more accurate estimation of the criterion of the enhancement and attenuation of homogeneous turbulence.

Numerical simulations have been playing important roles in the investigation of this complex phenomenon with many control parameters. \citet{Elghobashi-1993} and \citet{Elghobashi-1994} conducted numerical simulations of particulate turbulence. Their simulations were conducted with point-wise particles that obey the \citet{Maxey-1983} equation, and they showed the importance of the normalized particle velocity relaxation time (i.e.~the Stokes number). Although continuum approaches \citep{Crowe-1996} were also used, we have to resolve flow around each particle to accurately treat fluid-particle interactions. Numerical methods for such direct numerical simulations (DNS) with finite-size particles were proposed in this century \citep{Kajishima-2001,Ten-Cate-2004,Uhlmann-2005,Burton-2005}. For example, \citet{Kajishima-2001} numerically demonstrated turbulence enhancement by finite-size particles. Since then, numerical schemes \citep{Maxey-2017} have been developing to more easily and accurately treat the non-slip boundary condition on particles' surface. Thanks to these developments, many authors recently conducted DNS of particulate turbulence under realistic boundary conditions: for example, channel flow \citep{Uhlmann-2008,Shao-2012,Picano-2015,Wang-2016,Fornari-2016,Costa-2016,Costa-2018,Peng-2019ch}, pipe flow \citep{Peng-2019cy}, duct flow \citep{Lin-2017} and Couette flow \citep{Wang-2017}.

In the present study, as a first step towards the complete clarification, prediction and control of the interaction between solid particles and turbulence, we examine the simplest case: namely, the modulation of turbulence by finite-size solid spherical particles in a periodic cube. Many authors \citep{Ten-Cate-2004,Yeo-2010,Homann-2010,Lucci-2010,Lucci-2011,Gao-2013,Wang-2014,Uhlmann-2017,Schneiders-2017} numerically studied behaviors of finite-size particles in periodic turbulence. Concerning turbulent modulation, \citet{Ten-Cate-2004} conducted DNS of forced turbulence of particle suspension to demonstrate the enhancement of energy dissipation due to the excitation of particle-size flow. In particular, they showed that the energy spectrum was enhanced for wavenumber $k$ larger than the pivot wavenumber $k_p\approx0.72k_d$ with $k_d=2\pi/D$ being the wavenumber corresponding to the particle diameter $D$, whereas it was attenuated for $k<k_p$. Similar modulation of the energy spectrum was also observed by \cite{Yeo-2010}, \cite{Gao-2013} and \cite{Wang-2014}. An important observation in these studies is that the pivot wavenumber $k_p$ is approximately proportional to $k_d$ in forced turbulence \citep{Ten-Cate-2004,Yeo-2010}, though $k_d/k_p$ varies in decaying turbulence \citep{Gao-2013}. The importance of particle size was also emphasized by \citet{Lucci-2010,Lucci-2011}. More concretely, \cite{Lucci-2011} numerically demonstrated that the decay rate of turbulence depended on the particle size even if the Stokes number was identical. \citet{Gao-2013} demonstrated similar results, although they also emphasized the impact of the Stokes number on the turbulence modulation. Recall that once we fix the flow conditions, turbulence modulation can depend on, in addition to the number of particles, both the particle size and the Stokes number. Although the importance of the particle size is evident, the role of the Stokes number is still ambiguous. In particular, the condition for the turbulence modulation (i.e.~attenuation or enhancement) has not been explicitly described in terms of these particle properties because of the lack of systematic parametric studies. Besides, it is also desirable to predict the degree of turbulence modulation under given flow conditions and particle properties.

The present study aims at showing the condition for finite-size particles to attenuate turbulence in a periodic cube. To this end, we conduct a systematic parametric study by means of DNS of forced turbulence, and investigate turbulence modulation due to spherical solid particles with different diameters and Stokes numbers for a fixed volume fraction. Then, based on the obtained numerical results, we propose formulae that give the condition and degree of turbulence attenuation.

\section{Direct numerical simulations}
\subsection{Numerical methods}

The fluid velocity $\bm{u}(\bm{x},t)$ at position $\bm{x}$ and time $t$ is governed by the Navier--Stokes equation,
\begin{equation}
 \label{eq:ns}
 \frac{\partial \bm{u}}{\partial t}
 +
 \bm{u}\cdot\bm{\nabla u}
 =
 -\frac{1}{\rho_f}\:\bm{\nabla}p
 +
 \nu\nabla^2\bm{u}+\bm{f}
 +
 \bm{f}^{\leftarrow p}
 \:,
\end{equation}
and the continuity equation,
\begin{equation}
 \label{eq:divu=0}
 \bm{\nabla}\cdot\bm{u}=0
 \:,
\end{equation}
for an incompressible fluid in a periodic cube with side $L_0$ ($=2\pi$). Here, $p(\bm{x},t)$ is the pressure field, and $\rho_f$ and $\nu$ denote the fluid mass density and kinematic viscosity, respectively. In (\ref{eq:ns}), $\bm{f}^{\leftarrow p}(\bm{x},t)$ is the force due to suspended solid spherical particles, whereas $\bm{f}(\bm{x},t)$ is an external body force driving turbulence. In the present study, we examine the two cases with different kinds of external force $\bm{f}(\bm{x},t)$. One is a time-independent forcing \citep{Goto-2017}:
\begin{equation}
\label{eq:f^v}
 \bm{f}^{(v)}
 =
 (-\sin x\cos y, +\cos x\sin y, 0)
 \:.
\end{equation}
The other forcing $\bm{f}^{(i)}(\bm{x},t)$ is a force which keeps the energy input rate $P$ constant \citep{Lamorgese-2005}. This forcing is concretely expressed in terms of its Fourier transform $\widehat{\bm{f}^{(i)}}(\bm{k},t)$, where $\bm{k}$ is the wavenumber, as
\begin{equation}
 \label{eq:f^i}
 \widehat{\bm{f}^{(i)}}(\bm{k},t) 
 = 
 \left\{
 \begin{split}
  &\frac{P}{2E_f(t)}\:\widehat{\bm{u}}(\bm{k},t) &\quad &\text{if} \ \ 0<|\bm{k}| \leq k_f, \\
  &0 &\quad &\text{otherwise}.
 \end{split} \right. 
\end{equation}
In (\ref{eq:f^i}), $\widehat{\bm{u}}(\bm{k},t)$ and $E_f$ are the Fourier transform of $\bm{u}(\bm{x},t)$ and the kinetic energy 
\begin{equation}
 E_f
 = 
 \sum_{0<|\bm{k}|\leq k_f}\frac12\:|\widehat{\bm{u}}|^2
\end{equation}
in the forcing wavenumber range ($0<|\bm{k}|\leq k_f$), respectively. In (\ref{eq:f^i}), $P$ is arbitrary because the Reynolds number can be changed by changing $\nu$. We use the value $P=1$, whereas we set $k_f=1.5$ so that we can make the inertial range as wide as possible. Note that $\bm{f}^{(i)}$ sustains statistically homogeneous isotropic turbulence, whereas $\bm{f}^{(v)}$ sustains turbulence with a mean flow which is composed of four columnar vortices \citep{Goto-2017}.

In the present DNS, we use the second-order central finite difference on a staggered grid to estimate the spatial derivatives in (\ref{eq:ns}). We use $N^3=256^3$ grid points for the main series of DNS, and $512^3$ points for accuracy verifications. We summarize, in table \ref{t:para:turb}, other numerical parameters and the statistics of the single-phase turbulence. In the present study, we estimate the integral length $L(t)$ by $3\pi\int_0^\infty k'^{-1}E(k',t)\text{d}k'\big/4\int_0^\infty E(k',t)\text{d}k'$, where $E(k,t)$ is the energy spectrum, and the Taylor length $\lambda(t)$ by $\sqrt{10\nu K'(t)/\epsilon(t)}$, where $\epsilon(t)$ is the spatial average of the energy dissipation rate and $K'(t)$ is the turbulent kinetic energy:
\begin{equation}
\label{eq:K'}
 K'(t)
 =
 \frac12\:\Big\langle\Big|\bm{u}(\bm{x},t)-\bm{U}(\bm{x})\Big|^2\Big\rangle
 \qquad\text{with}\qquad
 \bm{U}(\bm{x})=\overline{\bm{u}(\bm{x},t)}
 \:.
\end{equation}
Here, $\langle\cdot\rangle$ and $\overline{\:\cdot\:}$ denote the spatial and temporal averages, respectively. Then, the Taylor-length-based Reynolds number is evaluated by $R_\lambda(t)=u'(t)\lambda(t)/\nu$, where $u'(t)=\sqrt{2K'(t)/3}$. We also estimate the Kolmogorov length by $\eta(t)=\epsilon(t)^{-\frac14}\nu^{\frac34}$. We have confirmed that the statistics shown in table \ref{t:para:turb} are common in Runs 256v and 512v, implying that the spatial resolution for the former run is fine enough.

We estimate the particle-fluid interaction force $\bm{f}^{\leftarrow p}(\bm{x},t)$ in (\ref{eq:ns}) by an immersed boundary method \citep{Uhlmann-2005}. In this method, we uniformly distribute $N_L$ Lagrangian force points on each particle surfaces \citep{Saff-1997,Lucci-2010} to estimate the interaction force $\widetilde{\bm{f}}^{\leftarrow p}$ by imposing the non-slip boundary condition of the fluid velocity on these points. The force $\bm{f}^{\leftarrow p}(\bm{x},t)$ is determined by redistributing $\widetilde{\bm{f}}^{\leftarrow p}$ onto grid points, whereas the force $\bm{f}^{\leftarrow f}_j$ and moment $\bm{L}^{\leftarrow f}_j$ around the particle center acting on the $j$th particle are estimated by integrating the reaction, $-\widetilde{\bm{f}}^{\leftarrow p}$, on the particle's surface. Then, we obtain the position $\bm{x}_j(t)$, velocity $\bm{v}_j(t)=\text{d}\bm{x}_j/\text{d}t$ and angular velocity $\bm{\omega}_j(t)$ of the $j$th particle ($1\leq j\leq N_p$ with $N_p$ being the number of particles) by integrating Newton's equations of motion:
\begin{equation}
 m\:
 \frac{\text{d}\bm{v}_j}{\text{d}t}
 =
 \bm{f}^{\leftarrow f}_j
 + 
 \bm{f}^{\leftrightarrow p}_j
 \label{eq:x_j}
\end{equation}
and
\begin{equation}
 I\:
 \frac{\text{d}\bm{\omega}_j}{\text{d}t}
 =
 \bm{L}^{\leftarrow f}_j
 \label{eq:o_j}
 \:.
\end{equation}
Here, we denote the diameter and mass density of the particles by $D$ and $\rho_p$, and therefore the mass and inertial moment of a particle are $m=\pi\rho_p D^3/6$ and $I=m D^2/10$, respectively. In (\ref{eq:x_j}), $\bm{f}^{\leftrightarrow p}$ is the interaction force between particles. For this, we only consider the normal component of the contact force due to the elastic collision, which is estimated by the standard discrete element method. For the estimation of $\bm{f}^{\leftrightarrow p}$, we neglect the frictional force and the lubrication effect. We have also neglected the gravity.

We numerically integrate (\ref{eq:ns}), (\ref{eq:x_j}) and (\ref{eq:o_j}) by the fractional step method \citep{Uhlmann-2005}, where we use the second-order Adams-Bashforth method instead of the three-step Runge-Kutta method. We also use a modified version \citep{Kempe-2012} of Uhlmann's immersed boundary method for particles with the smallest Stokes numbers in each Run (see table \ref{t:para} in the next subsection) when we integrate (\ref{eq:x_j}) and (\ref{eq:o_j}); it improves the numerical stability by modifying the evaluation method of $\bm{f}_j^{\leftarrow f}$ and $\bm{L}_j^{\leftarrow f}$ in these equations. We integrate the viscous term in (\ref{eq:ns}) by the second-order Crank-Nicolson method and the elastic force in (\ref{eq:x_j}) by the first-order Euler method. The discretized forms of the Poisson equation for the pseudo-pressure and the Helmholtz equation for the implicit integration of the viscous term are solved by the direct method with the fast Fourier transform (FFT). We also use FFT to estimate the body force $\bm{f}^{(i)}$ by (\ref{eq:f^i}), where we do not use any special treatment for the velocity at the grid points inside particles, since the particle size (see table \ref{t:para} in the next subsection) is always smaller than the forcing scale, $2\pi/k_f$. Our DNS codes have been validated by the test of a sedimenting sphere demonstrated in \S~5.3.1 of \cite{Uhlmann-2005}.

\begin{table}
\begin{center}
\begin{tabular}{lcccccccccc}
& $\bm{f}$ & $N^3$ & $\nu$ & $\overline{R_\lambda}$ & $L_0/\overline{L}$ & $\overline{L}/\overline{\eta}$ & $\overline{\eta}/\Delta x$ & CFL number \\
Run 256v & 
\multirow{2}{*}{$\bm{f}^{(v)}$} &
$256^3$ & 
\multirow{2}{*}{$8\times10^{-3}$} & 
\multirow{2}{*}{48} &
\multirow{2}{*}{5.3} &
\multirow{2}{*}{48} &
1.0 & 
\multirow{2}{*}{$6.4\times10^{-2}$} \\
Run 512v & 
&
$512^3$ & 
&
&
& 
& $2.0$ 
& \\
Run 256i & 
$\bm{f}^{(i)}$ &
$256^3$ & 
$7.13\times10^{-3}$ & 
94 & 
4.7 & 
54 &   
1.0 & 
$5.8\times10^{-2}$\\
\end{tabular}
\end{center}
\caption{
Parameters and statistics of single-phase turbulence.
$N^3$, number of grid points;
$L_0$ $(=2\pi)$, side of the numerical domain;
$\Delta x$ $(=L_0/N)$, grid width;
$\nu$, kinematic viscosity;
$R_\lambda$, Taylor-length-based Reynolds number;
$L$, integral length;
$\eta$, Kolmogorov length.
The CFL number is defined by the temporal average of 
$\sqrt{2K_t/3}\:\Delta t/\Delta x$ with the total kinetic energy $K_t$ per unit mass and the time increment $\Delta t$ of the temporal integration.}
\label{t:para:turb}
\end{table}

\begin{table}
\centering
($a$) Run 256v \vspace*{5pt}\\
\scalebox{0.85}{
\begin{tabular}{rrrrrrrrrrrrrrrrrr}
$D/\Delta x$ \ \ &
  8 &   8 &   8 &   8 &  8 &16 &  16 &  16 &  16 &  32 &  32 &  32 &  32 &  64 &  64 &  64 &  64  \\
$D/\overline{L}$ \ \ &
0.17&0.17 & 0.17 & 0.17 & 0.17 &0.33 & 0.33 & 0.33 & 0.33 & 0.66 & 0.66 & 0.66 & 0.66 & 1.3 & 1.3 & 1.3 & 1.3  \\
$D/\overline{\eta}$ \ \ &
7.8 & 7.8 & 7.8 & 7.8 & 7.8 & 16 & 16 & 16 & 32 & 32 & 32 & 32 & 63 & 63 & 63 & 63  \\
$\gamma$ \ \ &
  2 &   8 &  32 & 128 & 512 &  2 &   8 &  32 & 128 &   2 &   8 &  32 & 128 &   2 &   8 &  32 & 128  \\
$St$ \ \ &
0.51& 2.0 & 8.1 & 32 & 130 &2.0 & 8.1 & 32 & 130 & 8.1 & 32 & 130 & 520 & 32 & 130 & 520 & 2100 \\
$N_p$ \ \ &
512 & 512 & 512 & 512 & 512 & 64 &  64 &  64 &  64 &   8 &   8 &   8 &   8 &   1 &   1 &   1 &   1  \\
$N_L$ \ \ &
202 & 202 & 202 & 202 & 202 & 805 & 805 & 805 & 805 &3218 &3218 &3218 &3218 &12869&12869&12869&12869 \\
$Re_p$ \ \ &
11 & 25 & 37 & 42 & 43 &
41 & 65 & 80 & 83 & 
113 & 124 & 130  & 132  & 
 - &  - & -  & - \\
\end{tabular}}\vspace*{5pt}\\
($b$) Run 512v \vspace*{5pt}\\
\scalebox{0.85}{
\begin{tabular}{rrrrrr}
$D/\Delta x$ \ \ &
  16 &   16 &   16 &   16 & 16 \\
$D/\overline{L}$ \ \ &
0.17&0.17 & 0.17 & 0.17 & 0.17\\
$D/\overline{\eta}$ \ \ &
7.8 & 7.8 & 7.8 & 7.8 & 7.8 \\
$\gamma$ \ \ &
  2 &   8 &  32 & 128 & 512\\
$St$ \ \ &
0.51& 2.0 & 8.1 & 32 & 130 \\
$N_p$ \ \ &
512 & 512 & 512 & 512 & 512\\
$N_L$ \ \ &
805 & 805 & 805 & 805 & 805
\end{tabular}}\vspace*{5pt}\\
($c$) Run 256i \vspace*{5pt}\\
\scalebox{0.85}{
\begin{tabular}{rrrrrrrrrrrrrrrrrr}
$D/\Delta x$ \ \ &
 8 &   8 &   8 &   8 & 8 & 16 &  16 &  16 &  16 &  32 &  32 &  32 &  32 &  64 &  64 &  64 &  64  \\
$D/\overline{L}$ \ \ &
0.15 & 0.15 & 0.15 & 0.15 & 0.15 & 0.29 & 0.29 & 0.29 & 0.29 & 0.59 & 0.59 & 0.59 & 0.59 & 1.2 & 1.2 & 1.2 & 1.2  \\
$D/\overline{\eta}$ \ \ &
8.0 & 8.0 & 8.0 & 8.0 & 8.0 & 16 & 16 & 16 & 16 & 32 & 32 & 32 & 32 & 64 & 64 & 64 & 64  \\
$\gamma$ \ \ &
  2 &   8 &  32 & 128 & 512 & 2 &   8 &  32 & 128 &   2 &   8 &  32 & 128 &   2 &   8 &  32 & 128  \\
$St$ \ \ &
0.64& 2.6 & 10 & 41 & 170 & 2.6 & 10 & 41 & 170 & 10 & 41 & 170 & 660 & 41 & 170 & 660 & 2700 \\
$N_p$ \ \ &
512 & 512 & 512 & 512 & 512 &  64 &  64 &  64 &  64 &   8 &   8 &   8 &   8 &   1 &   1 &   1 &   1  \\
$N_L$ \ \ &
202 & 202 & 202 & 202 & 202 & 805 & 805 & 805 & 805 &3218 &3218 &3218 &3218 &12869&12869&12869&12869 \\
$Re_p$ \ \ &
11 & 23 & 35 & 42 & 44  &
42 & 63 & 82 & 88 & 
112 & 129 & 140 & 145  & 
- & - & - & -  \\
\end{tabular}}
\caption{
Parameters of the particles:
$D$, diameter;
$\gamma$ $(=\rho_p/\rho_f)$, mass density ratio;
$St$, Stokes number defined by $T$;
$N_p$, the number of particles;
$N_L$, the number of force points on a particle.
We use the values of $\overline{L}$ and $T$ of the single-phase turbulence (table \ref{t:para:turb}). Note also that the volume fraction, $\Lambda$, is fixed to be $8.2\times10^{-3}$ in all the cases. The particle Reynolds number, (\ref{eq:Rep}), is also listed in the bottom row of ($a$) and ($c$). We do not show $Re_p$ for the largest particles because the relative velocity cannot be estimated.}
\label{t:para}
\end{table}

\subsection{Parameters}

For a given external forcing, the parameters of fluid phase are the kinematic viscosity $\nu$, the mass density $\rho_f$, a characteristic length (e.g.~the integral length $\overline{L}$ or the Taylor length $\overline{\lambda}$) and a characteristic velocity (e.g.~the root mean square $\overline{u'}$ of fluctuation velocity). The parameters of particles are, on the other hand, the diameter $D$, the mass density $\rho_p$ and the number $N_p$ of particles. Therefore, there are four independent non-dimensional parameters. Here, we adopt $\overline{R_\lambda}=\overline{u'\lambda/\nu}$, the volume fraction $\Lambda$ of the particles, the non-dimensional particle diameter $D/\overline{L}$ and the particle Stokes number $St=\tau_p/T$. Here,
\begin{equation}
 \label{eq:tau_p}
 \tau_p=\frac{\gamma D^2}{18\nu} \qquad\text{($\gamma=\rho_p/\rho_f$)}
\end{equation}
is the relaxation time of particle velocity and $T=\overline{L}/\overline{u'}$ is the turnover time of the largest eddies. We conduct three series of DNS with fixed $\overline{R_\lambda}$ and $\Lambda$ $(=8.2\times10^{-3})$ by changing $D/\overline{L}$ and $St$; see tables \ref{t:para:turb} and \ref{t:para}. In \S~\ref{s:results} and \S~\ref{s:dis} (see figures \ref{fig:dU} and \ref{fig:formula}), we also discuss results of supplemental DNS for the smallest particles in Runs 256v and 256i with a smaller volume fraction ($\Lambda=4.1\times10^{-3}$).

\section{Results}
\label{s:results}

\begin{figure}
\begin{center}
\includegraphics[bb=0 0 643 268,width=\textwidth]{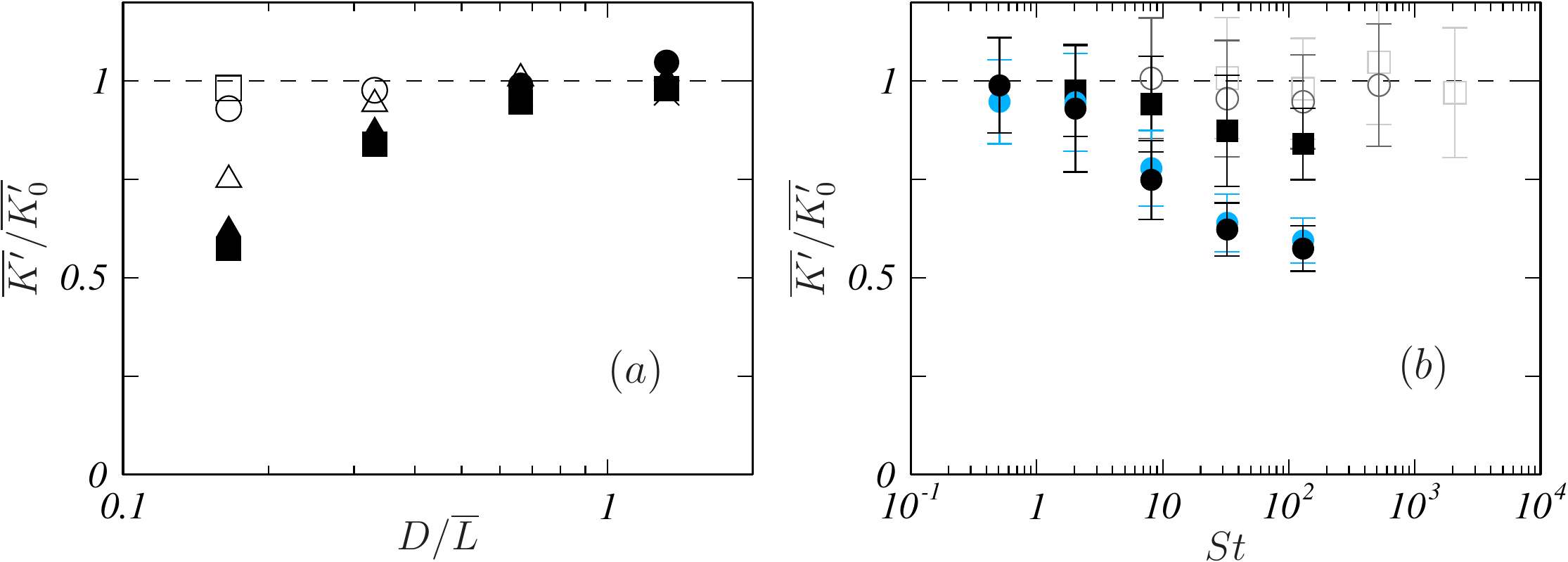} 
\end{center}
\caption{
($a$) Particle-size dependence of the temporal mean $\overline{K'}$ of the turbulent kinetic energy, which is normalized by the value $\overline{K'_0}$ for the single-phase flow. The results of Run 256v with forcing $\bm{f}^{(v)}$. Different symbols denote the results for different values of the Stokes number:
$St=0.51$, $\square$;
$2.0$, $\circ$;
$8.1$, $\triangle$;
$32$, $\blacktriangle$;
$130$, $\blacksquare$;
$520$, {\large$\bullet$};
$2100$, $\times$.
($b$) Stokes-number dependence of $\overline{K'}$. Different symbols correspond to different particle diameters:
$D/\overline{L}=0.17$, {\large$\bullet$};
$0.33$, $\blacksquare$;
$0.66$, $\circ$;
$1.3$, $\square$.
In the cases of the smallest particles ($D/\overline{L}=0.17$), we also show the results of higher-resolution DNS (Run 512v) with light-blue symbols. Error bars indicate the standard deviation of $K'(t)$.}
\label{fig:K'}
\begin{center}
\vspace*{10pt}
\includegraphics[bb=0 0 642 230,width=\textwidth]{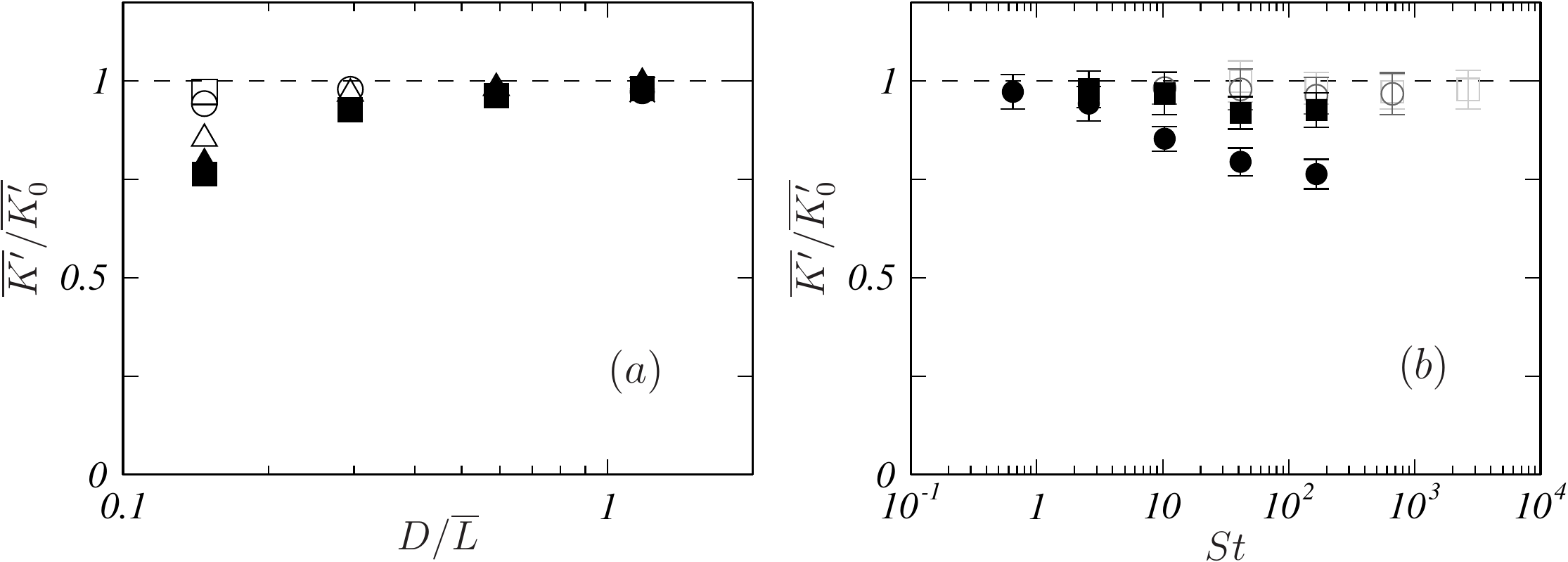}
\end{center}
\caption{Same as figure \ref{fig:K'} but for the other forcing $\bm{f}^{(i)}$ (Run 256i). ($a$) Different symbols denote the results with different values of the Stokes number:
$St=0.64$, $\square$;
$2.6$, $\circ$;
$10$, $\triangle$;
$41$, $\blacktriangle$;
$170$, $\blacksquare$;
$660$, $\bullet$;
$2700$, $\times$.
($b$) Different symbols correspond to different particle diameters:
$D/\overline{L}=0.15$, $\bullet$;
$0.29$, $\blacksquare$;
$0.59$, $\circ$;
$1.2$, $\square$.}
\label{fig:K'-i}
\end{figure}

The target of the present study is the attenuation of the turbulent kinetic energy defined by (\ref{eq:K'}). First, we examine the turbulence driven by the external force $\bm{f}^{(v)}$. We show the temporal average $\overline{K'}$, normalized by the value $\overline{K'_0}$ for the single-phase flow, in figure \ref{fig:K'}($a$) as a function of the particle diameter $D$ normalized by the integral length $\overline{L}$. Here, we compute the time average for the dulation of $250T$ in the statistically steady state. At the initial time, we distribute the particles uniformly on a three-dimensional lattice with vanishing velocity, and we exclude the transient period of about $19T$ before the system reaches the statistically steady state. On the other hand, we evaluate the spatial average $K'(t)$ of the turbulent kinetic energy of the fluid by using the method proposed by \cite{Kempe-2012} to calculate the volume fraction of the fluid phase in each grid cell. 

It is clear, in figure \ref{fig:K'}(a), that smaller particles are able to attenuate turbulence more significantly and no attenauation occurs when $D$ is as large as $\overline{L}$. This is consistent with the conventional view \citep{Gore-1989}. However, looking at the result with $D=0.17\overline{L}$ and $St=0.51$ for exmaple, it is also clear that $D\lesssim\overline{L}$ is not the sufficient condition for the attenuation and that the degree of the turbulence reduction depends on the Stokes number.

The $St$-dependence of the attenuation rate is evident in figure \ref{fig:K'}($b$). Looking at the case with the smallest particles $D=0.17\overline{L}$ ($\bullet$ in figure \ref{fig:K'}$b$), we can see that the attenuation is more significant for larger $St$ and it saturates for $St\gg1$, for which we observe about 43\% reduction of $\overline{K'}$. Recall that the volume fraction $\Lambda$ of the particles is only $8.2\times10^{-3}$. Although larger particles with $D=0.33\overline{L}$ ($\blacksquare$ in figure \ref{fig:K'}$b$) also attenuate the turbulence, the attenuation rate is smaller than the cases with $D=0.17\overline{L}$. However, the tendency that the attenuation rate, for fixed $D$, is larger for larger $St$ and it saturates for $St\gg1$ is common in the both cases with $D=0.17\overline{L}$ and $0.33\overline{L}$. Larger particles with $D=0.66\overline{L}$ or $1.3\overline{L}$ cannot attenuate turbulence even if $St\gg1$.

To verify the numerical accuracy, we also show the results of Run 512v in figure \ref{fig:K'}($b$). Recall that Runs 512v and 256v treat the common physical parameters (table \ref{t:para}) with different spatial resolutions for the smallest particles ($D=0.17\overline{L}$), since it is particularly important to show that the significant reduction of turbulence intensity with those small particles is not an artifact. It is therefore of importance to confirm that the results (light-blue symbols) with the higher resolution ($D/\Delta x=16$, Run 512v) and those (black ones) of Run 256v ($D/\Delta x=8$) are in good agreement. This validation of the numerical resolution is consistent with the previous study \citep{Uhlmann-2017} with the same immersed boundary method, which also used the resolution of $D/\Delta x=16$. Incidentally, the relatively large fluctuations indicated by error bars in figure \ref{fig:K'}($b$) do not imply large statistical errors, but they stem from the significant temporal fluctuations of turbulence driven by $\bm{f}^{(v)}$ \citep{Yasuda-2014,Goto-2017}.

Next, we look at the results (figure \ref{fig:K'-i}) with the other forcing $\bm{f}^{(i)}$. The trend of the attenuation of turbulence driven by $\bm{f}^{(i)}$ is similar to the case with $\bm{f}^{(v)}$ shown in figure \ref{fig:K'}; when $D\lesssim\overline{L}$, the turbulence intensity is attenuated more significantly when $St$ is larger (or $D$ is smaller) for fixed $D$ (or fixed $St$). We also notice that the attenuation rate of turbulence driven by $\bm{f}^{(i)}$ is smaller than the case with $\bm{f}^{(v)}$. This is due to the fact that there is no mean flow in turbulence driven by $\bm{f}^{(i)}$. We will discuss this difference below in more detail.

\begin{figure}
{\centering
\includegraphics[bb=0 0 642 501,width=\textwidth]{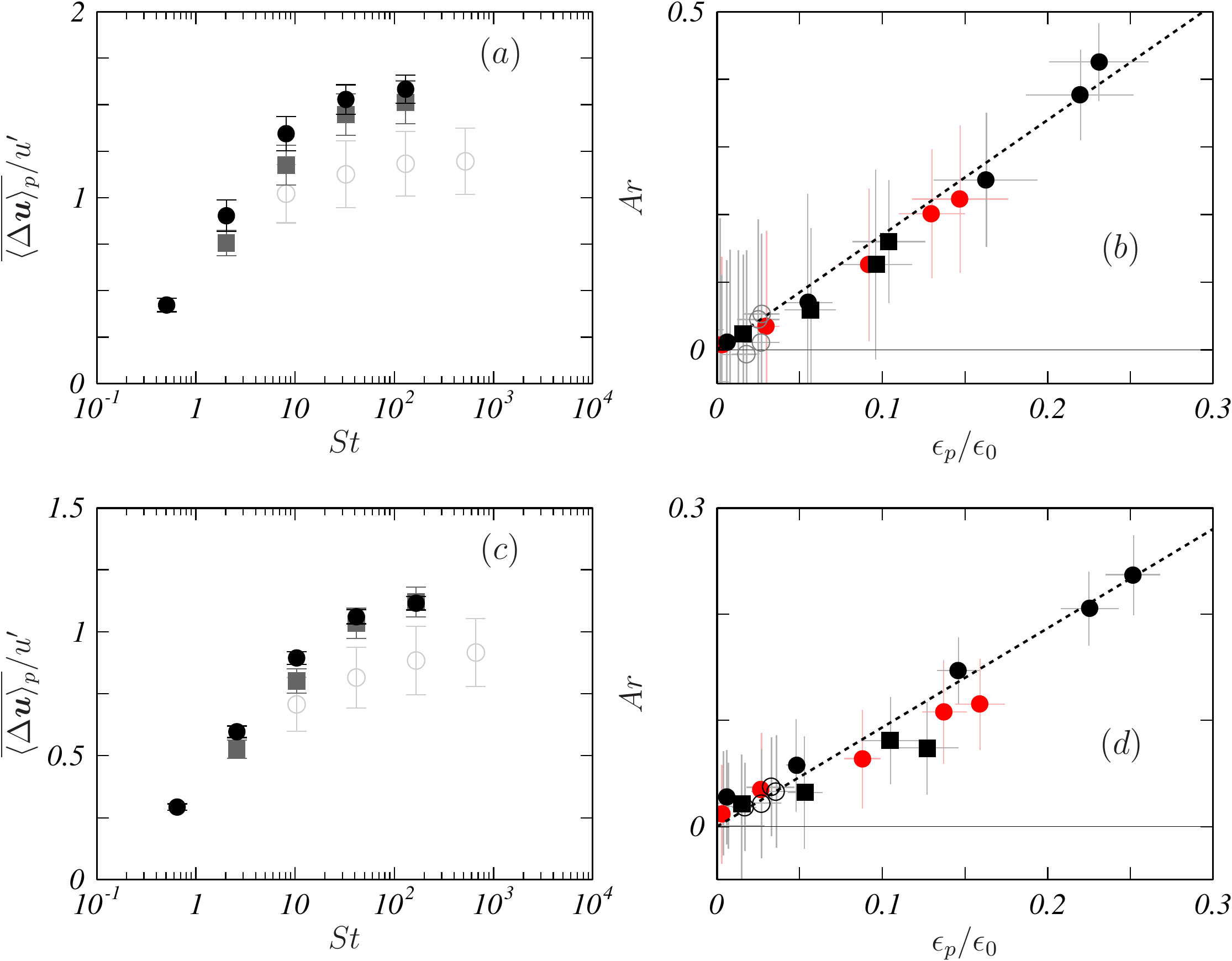}}
\caption{
($a$, $c$) Average relative velocity between a particle and the surrounding fluid. ($b$, $d$) Correlation between the attenuation rate (\ref{eq:Ar-def}) of the turbulent intensity and the estimate (\ref{eq:e_p}) of the energy dissipation rate $\epsilon_p$ due to particles normalized by the mean energy dissipation rate $\epsilon_0$ of single-phase turbulence. For the estimation of $\epsilon_p$, we put $C_p=1$. The results for ($a$, $b$) Run 256v (with $\bm{f}^{(v)}$) and ($c$, $d$) Run 256i (with $\bm{f}^{(i)}$). Different symbols are the results for different particle diameters:
($a$, $b$)
$D/\overline{L}=0.17$, $\bullet$;
$0.33$, $\blacksquare$;
$0.66$, $\circ$;
($c$, $d$)
$D/\overline{L}=0.15$, $\bullet$;
$0.29$, $\blacksquare$;
$0.59$, $\circ$.
In ($b$) and ($d$), red symbols represent results with a smaller volume fraction $\Lambda=4.1\times10^{-3}$ of the smallest particles [$D/\overline{L}=0.17$ in ($b$) and $0.15$ in ($d$)]; we show results for five cases of the mass ratio ($\gamma=2$, $8$, $32$, $128$ and $512$) in each panel. The proportional coefficients of the dotted lines in ($b$) and ($d$) are $1.7$ and $0.93$, respectively. Error bars indicate the standard deviations of the temporal fluctuations.}
\label{fig:dU}
\end{figure}

We have observed in figures \ref{fig:K'} and \ref{fig:K'-i} that, for fixed $D$, the attenuation is more significant for larger $St$ and it saturates when $St\gg1$. We can explain these observations by the facts that (i) the relative velocity magnitude between a particle and surrounding fluid is determined by $St$, and (ii) it is an increasing function of $St$ which tends to a value of $O(u')$ for $St\gg1$. To demonstrate these facts, we plot in figure \ref{fig:dU}($a$) the average relative velocity magnitude, $\overline{\langle|\Delta\bm{u}|\rangle_p}$, as a function of $St$ for Run 256v . Here, $\langle\cdot\rangle_p$ denotes the average over particles and we evaluate $\Delta\bm{u}$ for each particle by using the method proposed by \cite{Kidanemariam2013} and \cite{Uhlmann-2017}, where we define the velocity of the surrounding fluid of a particle by the average fluid velocity on the surface of the sphere with diameter $2D$ concentric with the particle. 

It is clear that the relative velocity magnitude tends to be a value of $O(u')$ when $St\gg1$ in the cases $D=0.17\overline{L}$ ($\bullet$) and $0.33\overline{L}$ ($\blacksquare$). Note that for larger particles (e.g.~the results shown in light gray for $D=0.66\overline{L}$) the estimated values of $\Delta\bm{u}$ may have less meaning. In particular, the estimated fluid velocity has no physical meaning when $D\gtrsim\overline{L}$ because it is the average of fluid velocity over a domain much larger than the largest eddies. This is the reason why we have excluded the data for the largerst particles ($D=1.3\overline{L}$) from figure \ref{fig:dU}(a) and the following arguments.

Similar dependence of $\overline{\langle|\Delta\bm{u}|\rangle_p}$ on $St$ and $D$ is observed in figure \ref{fig:dU}($c$) for the case (Run 256i) with the other forcing $\bm{f}^{(i)}$. Looking at the results with $D=0.15\overline{L}$ ($\bullet$) and $0.29\overline{L}$ ($\blacksquare$), we can see that the relative velocity magnitude is larger for larger $St$ and it tends to a value for $St\gg1$. It is clear in figures \ref{fig:dU}($a$) and \ref{fig:dU}($c$) that the velocity difference magnitude only weakly dependent on the particle size. This is reasonable because the Stokes number $St$ ($=\tau_p/T$) determines particles' ability to follow the swirling of the largest (i.e.~most energetic) eddies. We also notice that the relative velocity magnitude normalized by $u'$ is larger for $\bm{f}^{(v)}$ than $\bm{f}^{(i)}$. Since turbulence driven by $\bm{f}^{(v)}$ is accompanied by mean flow, the velocity of surrounding fluid, and therefore $|\Delta\bm{u}|$, can be larger.

\begin{figure}
\begin{center}
\includegraphics[bb=0 0 677 310,width=0.9\textwidth]{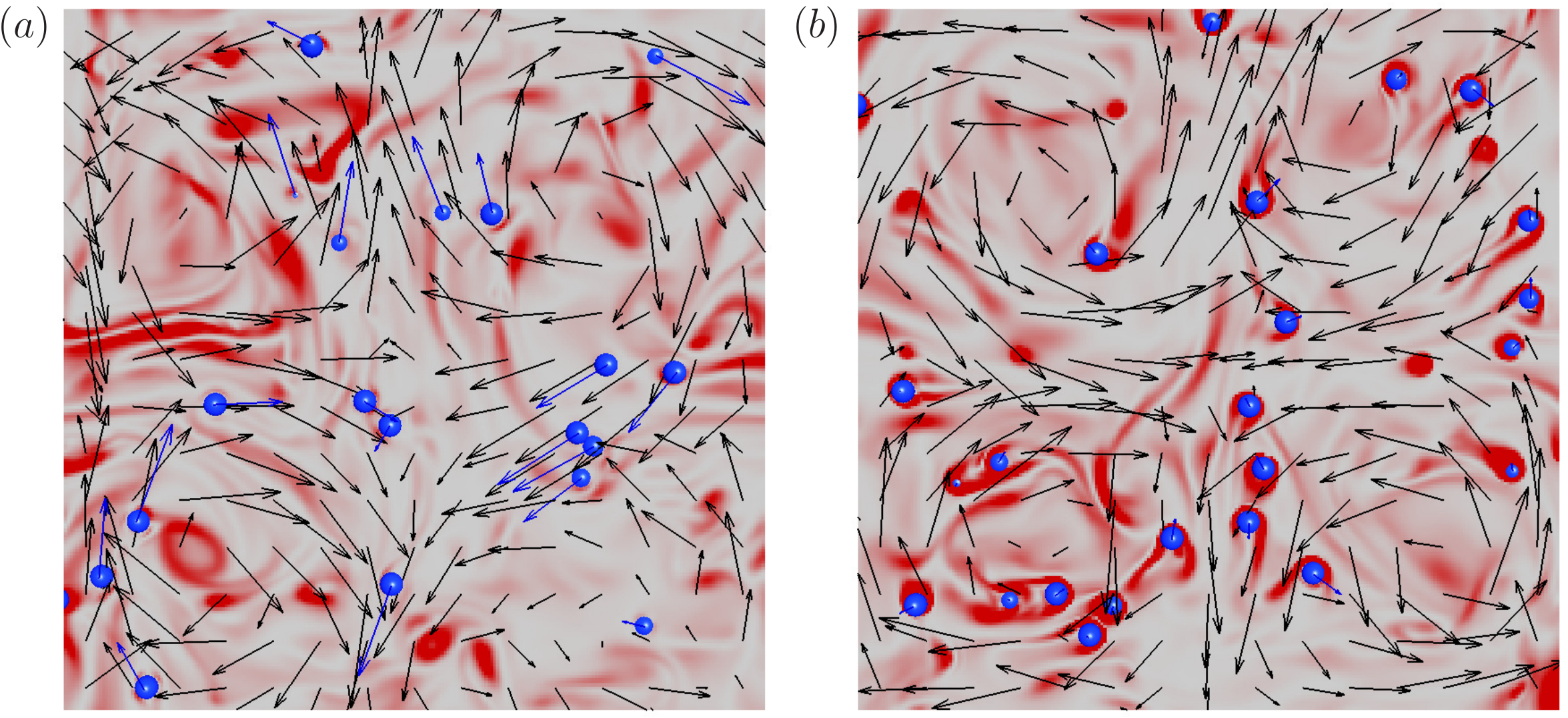} 
\end{center}
\caption{Visualization of flow and particle motions on the
$z=0$ plane for Run 256v. 
Blue balls, particles [$D=0.17\overline{L}$ ($=7.8\overline{\eta}$)]; 
blue arrows, particle velocity;
black arrows, fluid velocity;
background colour, enstrophy magnitude (redder colour
implies larger magnitudes). 
The Stokes number is ($a$)
$St=0.51$ and ($b$) 130. Supplemental movie is also available online.}
\label{fig:viz}
\end{figure}

We may also confirm the $St$-dependence of the relative velocity in visualizations. Figure \ref{fig:viz} shows snapshots of flow and particle motions on a crosssection $(z=0)$ for Run 256v. Black arrows show the flow, which is composed of four vortex columns sustained by $\bm{f}^{(v)}$, (\ref{eq:f^v}), whereas blue balls are the particles ($D=0.17\overline{L}$) with two different values of $St=0.51$ in ($a$) and $130$ in ($b$). Comparing the particle velocity (blue arrows) to the fluid velocity, we can see that the relative velocity is much more significant for the larger $St$. It is also remarkable that large enstrophy is produced in the wakes of the particles with larger $St$. As will be explained below, this large relative velocity and the resulting vortex shedding in large $St$ cases are the cause of the turbulence attenuation. 

Since we have computed the relative velocity, we can estimate the energy dissipation rate per unit mass due to the shedding vortices around particles by
\begin{equation}
\label{eq:e_p}
 \epsilon_p
 =
 C_p
 \Lambda\:
 \frac{\overline{\langle|\Delta\bm{u}|^3\rangle_p}}{D}
 \:.
\end{equation}
Here, $C_p$ is a constant and $\Lambda$ is the volume fraction of the particles. The estimation (\ref{eq:e_p}) of $\epsilon_p$ is derived under the assumption that the energy dissipation rate in the wake behind a single particle is balanced with the energy input rate $P_p$ due to the force from the particle to fluid. Since $P_p$ depends only on $D$ and $|\Delta\bm{u}|$ when the particle Reynolds number $Re_p$ [see (\ref{eq:Rep}), below] is large, the dimensional analysis leads to $P_p\sim|\Delta\bm{u}|^3/D$. Then, the mean energy dissipation rate due to all particles may be estimated by (\ref{eq:e_p}) with the factor of $\Lambda$ because the volume fraction of particle wakes is proportional to $\Lambda$. The estimation of $\epsilon_p$ by (\ref{eq:e_p}) is an approximation because, in a more precise sence, $C_p$ weakly depends on $Re_p$. This approximation is however sufficient in the following arguments. The additional energy dissipation rate $\epsilon_p$ is the key quantity to understanding the turbulence attenuation. More concretely, when the relative velocity is non-negligible, shedding vortices enhance turbulent fluctuating velocity at scales smaller than the particle size $D$. This enhancement was demonstrated in previous studies \citep{Ten-Cate-2004,Yeo-2010,Wang-2014} by investigating the energy spectrum. In particular, they showed that energy spectrum, $E(k)$, was enhanced (attenuated) for the wavenumbers $k$ larger (smaller) than the pivot wavenumber $k_p\approx0.6$--$0.9k_d$ with $k_d=2\pi/D$. In the present DNS, we may estimate $k_p$ in the case with the smallest particles because the other cases show only moderate attenuations. By estimating the energy spectrum without special treatments of the existence of particles, we observe that the smallest particles ($D\approx8\overline{\eta}$ in both cases of Runs 256v and 256i) attenuate $E(k)$ for $k\lesssim0.5k_d$ whereas strongly enhance it for $k\gtrsim k_d$ (figures are omitted). These observations are consistent with the proposed scenario of turbulence attenuation; that is, particles acquire their energy from the largest enegetic eddies and then bypass the energy cascading process to directly dissipate the energy at the rate $\epsilon_p$ in their wakes.

In fact, it is evident in figures \ref{fig:dU}($b$) and \ref{fig:dU}($d$) that the attenuation rate defined by 
\begin{equation}
\label{eq:Ar-def}
 Ar
 =
 \frac{\overline{K'_0}-\overline{K'}}{\overline{K'_0}}
\end{equation}
is approximately proportional to $\epsilon_p$. This is the most important observation of the present DNS. We also show in figure \ref{fig:dU}($b$, $d$) results (red symbols) with a smaller volume fraction ($\Lambda=4.1\times10^{-3}$) for the smallest particle cases. We can see that the relation between $Ar$ and $\epsilon_p$ is independent of $\Lambda$, which further verifies the estimation (\ref{eq:e_p}) of $\epsilon_p$. Note that the proportional coefficient, $Ar/(\epsilon_p/\epsilon_0)$, is about twice larger for the turbulence driven by $\bm{f}^{(v)}$ than that by $\bm{f}^{(i)}$. We will show, in the next section, the origin of this difference [see (\ref{eq:formula+})]. 

By using the estimated relative velocity magnitude, we can also estimate the particle Reynolds number
\begin{equation}
\label{eq:Rep}
 Re_p
 =
 \frac{D\overline{\langle|\Delta\bm{u}|\rangle_p}}{\nu}
\end{equation}
to see if $Re_p$ is large enough for vortex shedding. The estimated values are listed in table \ref{t:para}. For example, for $St=32$, 
$Re_p=42$ (for $D=0.17\overline{L}$), 
$80$ (for $0.33\overline{L}$) and
$124$ (for $0.66\overline{L}$). This means that vortices are shedding from the particles in these cases. It is, however, important to emphasize that although $Re_p\gtrsim1$ is a necessary condition for the turbulence attenuation, large $Re_p$ does not always imply a large attenuation rate, which depends on $D$.

\section{Discussions}
\label{s:dis}

On the basis of the DNS results shown in the previous section, we discuss the physical mechanism of turbulence attenuation in the present system. Figures \ref{fig:K'} and \ref{fig:K'-i} imply that turbulence can be attenuated more significantly by smaller particles, and no attenuation occurs when $D/\overline{L}\approx1$. Therefore, here we restrict ourselves to the cases of the attenuation by small particles; more precisely,
\begin{equation}
\label{eq:D<L}
 D\lesssim\overline{L}
 \:.
\end{equation}
It is also an important observation that vortex shedding from particles is enhanced when turbulence is significantly attenuated (figure \ref{fig:viz} and the supplemental movie). This implies that, when vortices are shed from particles smaller than $\overline{L}$, the intrinsic turbulent energy cascade is bypassed and the energy dissipation is enhanced by the shedding vortices, which leads to the attenuation. In the following subsections, we consider the condition and degree of the turbulence attenuation due to this mechanism.

\subsection{Condition for turbulence attenuation}

Let us derive the condition for turbulence attenuation. For simplicity, in this section, we neglect the temporal fluctuations of $L(t)$, $\lambda(t)$, $K'(t)$ and $u'(t)$ and omit the over-bars of $\overline{L}$, $\overline{\lambda}$, $\overline{K'}$ and $\overline{u'}$. The DNS results shown in the previous section (see figures \ref{fig:dU}$b$ and \ref{fig:dU}$d$) indicate that the attenuation rate is determined by the energy dissipation rate (\ref{eq:e_p}) due to shedding vortices. Therefore, turbulence attenuation requires the conditions for shedding vortices to acquire their energy from the turbulence: (i) there exists non-negligible (i.e.~$O(u')$) relative velocity between particles and their surrounding fluid and (ii) the particle Reynolds number (\ref{eq:Rep}) is large enough for shedding vortices.

First, we examine (i), which is the condition for particles not to follow the surrounding flow. In other words, the particle velocity relaxation time $\tau_p$ is larger than the turnover time of the largest eddies: i.e.~$St\gtrsim1$. Estimating $\tau_p$ by (\ref{eq:tau_p}), we can express this condition ($St\gtrsim1$) as
\begin{equation}
\label{eq:cnd-i}
 D
 \gtrsim
 \sqrt{\frac{18}{\gamma Re}}\:L
 \sim
 \frac{\lambda}{\sqrt{\gamma}}
 \:.
\end{equation}
Here, we have defined the Reynolds number by $Re=u'L/\nu$ and used $T=L/u'$ and the expression 
\begin{equation}
 \label{eq:Taylor}
 \epsilon=\frac{15\nu u'^2}{\lambda^2}
 \sim \frac{u'^3}{L}
\end{equation}
of the energy dissipation rate in isotropic turbulence \citep{Taylor-1935I}. 

Equation (\ref{eq:cnd-i}) implies that the sufficient velocity difference between particles and fluid requires that particle diameter $D$ must be larger than a length proportional to the Taylor length $\lambda$. Note however that, when the mass density ratio $\gamma$ is much larger than $1$, particles smaller than $\lambda$ can attenuate turbulence because of the coefficient $1/\sqrt{\gamma}$ on the right-hand side of (\ref{eq:cnd-i}). Indeed, this is the case for some parameters of the present DNS; for example, for Run 256v (see table \ref{t:para:turb}) although $D=0.17{L}$ is comparable with $\lambda$, $St$ can be much larger than $1$ when $\gamma\gg1$, and in such cases turbulence is significantly attenuated (figure \ref{fig:K'}).

Next, we examine the second condition (ii). When (\ref{eq:cnd-i}) holds, the relative velocity magnitude is $O(u')$ (figures \ref{fig:dU}$a$ and \ref{fig:dU}$c$), and therefore the particle Reynolds number (\ref{eq:Rep}) is $Re_p\approx u'D/\nu$. The condition for $Re_p$ to be larger than $O(1)$ is, therefore, expressed as
\begin{equation}
\label{eq:cnd-ii}
 D
 \gtrsim
 L/Re
 \:.
\end{equation}
For $Re\gg1$, if (\ref{eq:cnd-i}) holds, (\ref{eq:cnd-ii}) also holds. Hence, (\ref{eq:cnd-i}) gives the lower bound of $D$ for the turbulence attenuation by small particles.

\subsection{Estimation of attenuation rate}
\label{s:rate}

Further developing the above arguments, we may also estimate the attenuation rate of $K'$. Here, we assume that, if $D\ll L$, particles have only limited impact on the mean flow; this is indeed the case in the present system with mean flow driven by $\bm{f}^{(v)}$. Under this assumption, the energy input rate, $\langle\bm{U}\cdot\bm{f}^{(v)}\rangle$, is the same as in the single-phase turbulence. Hence, because of the statistical stationarity, the mean energy dissipation rate of the particulate turbulence is approximately equal to the value
\begin{equation}
 \label{eq:e0=}
 \epsilon_0=C_\epsilon\frac{(K_0+K'_0)^{3/2}}{L}
\end{equation}
for the single-phase flow. Here, $C_\epsilon$ ($=O(1)$) is a flow-dependent constant \citep{Goto-2009}, and $K_0$ and $K'_0$ denote the kinetic energy of the mean and fluctuating single-phase flow, respectively. Incidentally, in the turbulence driven by $\bm{f}^{(i)}$, although the mean flow is absent, the energy input rate is the same, by construction (\ref{eq:f^i}) of the forcing, in the single-phase and particulate flows.

\begin{figure}
\centering \includegraphics[bb=0 0 355 261,width=0.6\textwidth]{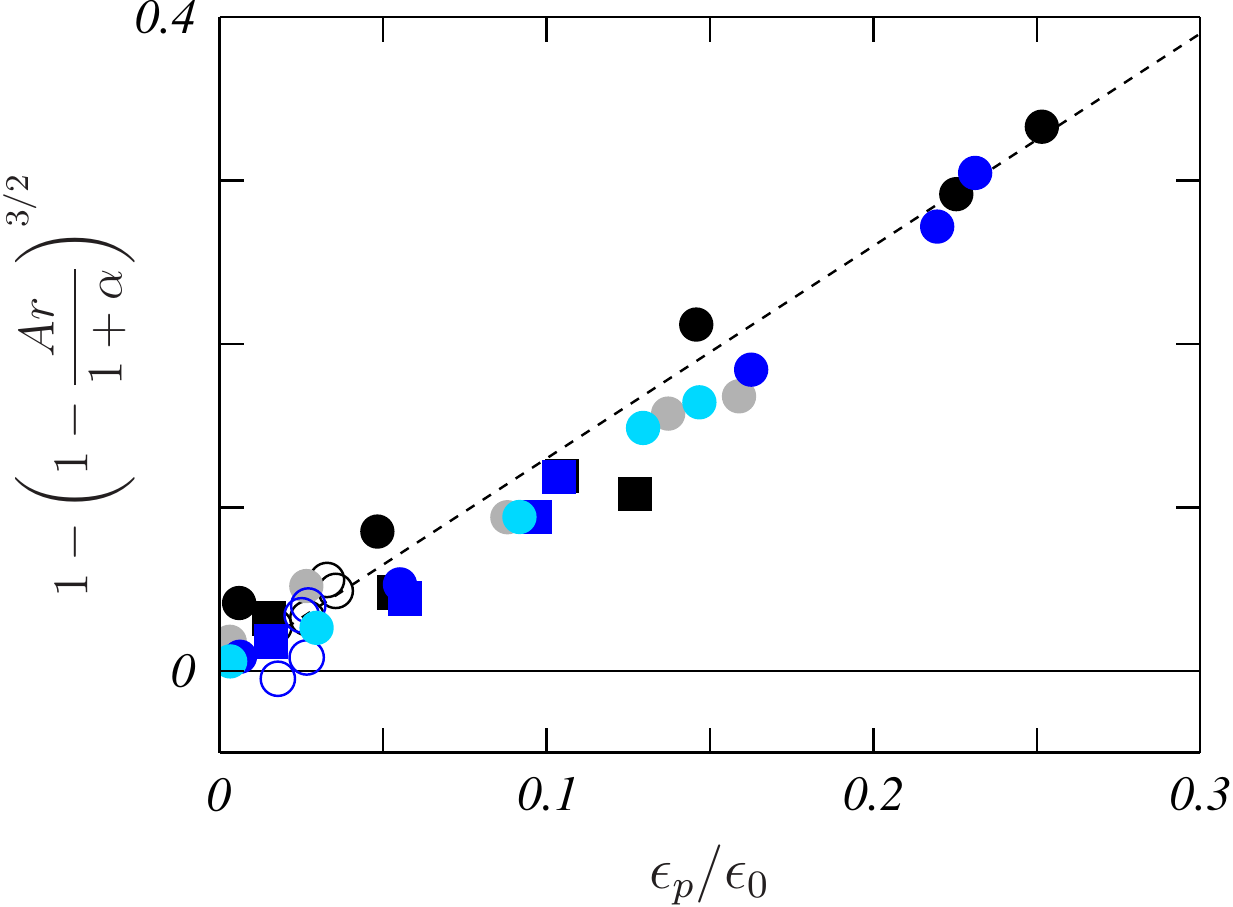}
\caption{Verification of (\ref{eq:formula}), according to which we replot the data in figures \ref{fig:dU}($b$) and \ref{fig:dU}($d$) with blue and black symbols, respectively. Darker and lighter symbols denote the cases with $\Lambda=8.2\times10^{-3}$ and $4.1\times10^{-3}$, respectively. The dotted line indicates $1.3\epsilon_p/\epsilon_0$.}
\label{fig:formula}
\end{figure}

In particulate turbulence with a small volume fraction of particles, the inputted energy is either transfered to the Kolmogorov scale by the energy cascading process from the forcing-scale eddies or dissipated in the wake behind particles. Hence, the energy dissipation rate is the sum of $\epsilon_c$ through the energy cascade and $\epsilon_p$ in the wake of the particles (i.e.~the energy dissipation rate bypassing the energy cascade):
\begin{equation}
 \label{eq:e0=ec+ep}
 \epsilon_0=\epsilon_c+\epsilon_p
 \:.
\end{equation}
Here, $\epsilon_c$ is expressed by
\begin{equation}
 \label{eq:ec=}
 \epsilon_c=C_\epsilon\frac{(K_0+K')^{3/2}}{L}
\end{equation}
in terms of the modulated turbulent kinetic energy $K'$. Then, substituting (\ref{eq:e0=}) and (\ref{eq:ec=}) into (\ref{eq:e0=ec+ep}) divided by $\epsilon_0$, we obtain the formula 
\begin{equation}
 \label{eq:formula}
 1-\left(1-\frac{Ar}{1+\alpha}\right)^{3/2}=\frac{\epsilon_p}{\epsilon_0}
\end{equation}
for the attenuation rate $Ar$ defined by (\ref{eq:Ar-def}). In (\ref{eq:formula}), $\alpha$ denotes the ratio
\begin{equation}
 \alpha=\frac{K_0}{K'_0}
\end{equation}
%
%
between the mean and fluctuation energy of single-phase flow: $\alpha=0$ for the turbulence driven by $\bm{f}^{(i)}$, whereas $\alpha$ is numerically estimated as $1.86/1.89\approx0.98$ for $\bm{f}^{(v)}$. Note that although $C_\epsilon$ in (\ref{eq:e0=}) and (\ref{eq:ec=}) depends on flow, (\ref{eq:formula}) is independent of $C_\epsilon$. This means that the formula (\ref{eq:formula}) is flow-independent. In fact, by using (\ref{eq:formula}), the two data sets of $Ar$ in figures \ref{fig:dU}($b$) and \ref{fig:dU}($d$) for the two kinds of forcing collapse (figure \ref{fig:formula}). The formula (\ref{eq:formula}) further reduces to 
\begin{equation}
 \label{eq:formula+}
 Ar\sim\frac{(1+\alpha)\:\epsilon_p}{\epsilon_0}
\end{equation}
when $Ar$ is not too large. This explains the reason why the proportional constant, $Ar/(\epsilon_p/\epsilon_0)$, figure \ref{fig:dU}($b$), is approximately twice larger than that in figure \ref{fig:dU}($d$). Recall that $\alpha+1\approx1.98$ for $\bm{f}^{(v)}$ and $\alpha+1=1$ for $\bm{f}^{(i)}$.

We emphasize that (\ref{eq:formula}) can predict the attenuation rate $Ar$, if we know $\epsilon_p$. By using (\ref{eq:e_p}), we may estimate $\epsilon_p$ for $St\gg1$ because $|\Delta\bm{u}|=cu'$ for $St\gg1$ with a flow-dependent constant $c$ (figures \ref{fig:dU}$a$ and \ref{fig:dU}$c$). Then, we may rewrite (\ref{eq:formula}) as
\begin{equation}
 1-\left(1-\frac{Ar}{1+\alpha}\right)^{3/2}=\frac{C_p'\Lambda L}{D}
 \qquad (St\gg1)
 \:.
\end{equation}
Here, $C_p'\sim c^3C_p/C_\epsilon$ is also a flow-dependent constant. The above equation may reduce to
\begin{equation}
 \label{eq:formula++}
 Ar\sim\frac{(1+\alpha)\Lambda L}{D}
 \qquad (St\gg1)
\end{equation}
when $Ar$ is not too large. This simple expression (\ref{eq:formula++}) means that the attenuation due to the considered mechanism occurs when (\ref{eq:D<L}) holds with a sufficient volume fraction $\Lambda$. In other words, the upper bound of the attenuation by small particles is given by (\ref{eq:D<L}). It also explains that the attenuation rate $Ar$ is larger for smaller $D$. Hence, combining this with the condition (\ref{eq:cnd-i}), we conclude that, for fixed $\Lambda$ and $\gamma$, particles with the size proportional to $\lambda/\sqrt{\gamma}$ most effectively attenuate turbulence intensity. Since the numerical verification of this conclusion requires DNS with further smaller particles, we leave it for future studies. It is also worth mentioning that $\epsilon_p$ [see (\ref{eq:e_p}) and figure \ref{fig:dU}($b$, $d$)], and therefore $Ar$ approximated by (\ref{eq:formula++}), are proportional to the volume fraction $\Lambda$. This explains the reason why larger mass fraction ($\gamma\Lambda$) generally tends to lead larger turbulence attenuation because $St$ is larger for larger $\gamma$.

\section{Conclusions}

We have derived the conditions (\ref{eq:D<L}) and (\ref{eq:cnd-i}), i.e.~$\overline{L}\gtrsim D \gtrsim\overline{\lambda}/\sqrt{\gamma}$, for the dilute additives of solid spherical particles, without the gravity, to attenuate turbulence in a periodic cube. First, we have numerically verified the conventional picture that the attenuation is due to the additional energy dissipation rate $\epsilon_p$, (\ref{eq:e_p}), caused by shedding vortices around  particles; more concretely, we have shown in figures \ref{fig:dU}($b$) and \ref{fig:dU}($d$) that the attenuation rate $Ar$ is approximately proportional to $\epsilon_p$. This result immediately leads to the attenuation condition because the attenuation occurs when $\epsilon_p$, (\ref{eq:e_p}), takes a finite value, which requires a finite relative velocity $|\Delta\bm{u}|$ between particles and their surrounding fluid; i.e.~$St\gtrsim1$. In fact, as shown in figures \ref{fig:dU}($a$) and \ref{fig:dU}($c$), $|\Delta\bm{u}|$ takes finite values when $St\gtrsim1$ and it tends to a value of $O(u')$ for $St\gg1$. The condition, $St\gtrsim1$, leads to (\ref{eq:cnd-i}) for the particle diameter $D$; and if (\ref{eq:cnd-i}) holds, then $Re_p\gg1$ also holds and therefore vortices are shedding from the particles. Hence, (\ref{eq:cnd-i}) gives the lower bound of $D$ for the turbulence attenuation. In other words, since particles smaller than $\overline{\lambda}/\sqrt{\gamma}$ behave like tracers for the largest energetic eddies, they cannot modulate them. 

The picture of the turbulence attenuation due to the shedding vortices also leads to the estimation of the attenuation rate. The simple argument developed in \S~\ref{s:rate} leads to (\ref{eq:formula}), which well explains the DNS results (figure \ref{fig:formula}). We emphasize that (\ref{eq:formula}) is a formula independent of forcing schemes. For $St\gg1$, (\ref{eq:formula}) reduces to $Ar\sim{\Lambda\overline{L}}/{D}$, (\ref{eq:formula++}), which implies that, for a given volume fraction, smaller particles which satisfy (\ref{eq:cnd-i}) more effectively attenuate turbulence. This is consistent with the DNS results (figures \ref{fig:K'} and \ref{fig:K'-i}). Hence, in turbulence at sufficiently high Reynolds numbers (and therefore $\overline{L}\gg\overline{\lambda}\gg\overline{\eta}$), particles with a size proportional to $\overline{\lambda}/\sqrt{\gamma}$ most significantly attenuate turbulence under the condition that the Reynolds number $Re$, the mass ratio $\gamma$ and the volume fraction $\Lambda$ are fixed. Furthermore, since $Ar$ is proportional to $\Lambda\overline{L}/D$ for $St\gg1$, turbulence is hardly attenuated when $D$ is as large as $\overline{L}$ (see also figures \ref{fig:K'} and \ref{fig:K'-i}). Therefore, (\ref{eq:D<L}) gives the upper bound of $D$ for the attenuation by the considered mechanism.

Recall that we have only considered turbulence attenuation by small particles. Although it is difficult to conduct DNS with particles larger than $\overline{L}$ in turbulence at similar Reynolds numbers, we may expect only small relative velocity for $D\gtrsim\overline{L}$ in the present system, where neither gravity nor mean flow larger than $\overline{L}$ exist. Then, vortices are not shed from such large particles. Incidentally, when the mean-flow or gravitational effects are important, the relative velocity between particles and fluid creates vortices, which can lead to turbulence modulation.

Before closing this article, it is worth mentioning the possibility that particles can modulate turbulence even if they do not satisfy (\ref{eq:cnd-i}) because they can interrupt energy cascade in the inertial range. More concretely, if particles' velocity relaxation time $\tau_p$ is comparable with the turnover time $\tau(\ell)$ of eddies with size $\ell$ in the inertial range, they follow the motion of eddies larger than $\ell$, but they have relative velocity with those smaller than $\ell$. Since larger eddies have more energy, the relative velocity between particles and fluid is determined by the eddies with size $\ell$. Therefore, the particles acquire their energy from such eddies with size $\ell$ and the some part of cascading energy at scales smaller than $\ell$ may be bypassed by the shedding vortices behind particles and dissipated in the wake of particles. Such a phenomenon is to be numerically observed in turbulence at higher Reynolds numbers in the near future.

\backsection[Acknowledgements]{The DNS were conducted under the supports of the NIFS Collaboration Research Program (20KNSS145) and under the supercomputer Fugaku provided by RIKEN through the HPCI System Research projects (hp210207). SG thanks the late professor Michio Nishioka for relevant discussions in the laboratory.}

\backsection[Funding]{This study was partly supported by JSPS Grant-in-Aids for Scientific Research (20H02068).}

\backsection[Declaration of interests]{The authors report no conflict of interest.}

\backsection[Author ORCIDs]{\\
Sunao Oka \url{https://orcid.org/0000-0001-6398-9213} \\
Susumu Goto \url{https://orcid.org/0000-0001-7013-7967}.}


\end{document}